\documentclass[a4paper,11pt]{article}

\usepackage{jheppub} 

\usepackage[T1]{fontenc} 

\usepackage{amsmath,amsfonts,amssymb,amsthm}
\usepackage{hyperref}
\hypersetup{
    pdfstartview=FitH,
    bookmarksnumbered=true,
    colorlinks,
    breaklinks,
    urlcolor=RoyalBlue,
    linkcolor=RoyalBlue,
    citecolor=NavyBlue
}
\usepackage{graphicx}
\DeclareGraphicsExtensions{.pdf,.png,.jpg,.mps}
\usepackage{booktabs}
\usepackage[dvipsnames]{xcolor}
\usepackage{tikz}
\usetikzlibrary{calc,matrix,decorations.pathmorphing,decorations.markings,arrows,positioning,intersections,mindmap,backgrounds}
\usepackage{cancel}

\usepackage{amsmath,amsthm,amsfonts,amssymb,amscd,mathtools
}
\usepackage{physics}
\usepackage{diagbox}
\usepackage{pifont}

\newcommand{\zb}{\bar{z}}
\newcommand{\hb}{\bar{h}}

\title{\boldmath Graviton Scattering in $\mathrm{AdS}_5\times\mathrm{S}^5$ at Two Loops}

\author{Zhongjie Huang}
\author{and Ellis Ye Yuan}
\affiliation{Zhejiang Institute of Modern Physics, Department of Physics, Zhejiang University,\\866 Yuhangtang Road, Hangzhou, Zhejiang 310058, China}

\affiliation{Joint Center for Quanta-to-Cosmos Physics, Zhejiang University,\\866 Yuhangtang Road, Hangzhou, Zhejiang 310058, China}

\emailAdd{zjhuang@zju.edu.cn}
\emailAdd{eyyuan@zju.edu.cn}

\abstract{We report a result for the third-order correction in $1/N$ expansion to the four-point correlator of the stress tensor multiplet in $\mathcal{N}=4$ super Yang--Mills theory at large 't Hooft coupling, which corresponds to the two-loop scattering of four gravitons in the dual $\mathrm{AdS}_5\times\mathrm{S}^5$ supergravity. This is obtained by bootstrapping an educated ansatz based on intuitions from the hidden 10-dimensional conformal symmetry.}

\begin{document} 
\maketitle
\flushbottom

\section{Introduction}

Correlators of chiral primary operators (CPO) are crucial for understanding the correspondence between $\mathcal{N}=4$ super Yang--Mills with gauge group $\mathrm{SU}(N)$ (SYM) and its AdS dual \footnote{See \cite{Heslop:2022qgf} for a nice review on these correlators.}. In the supergravity limit they serve as useful probe for particle dynamics in the bulk. The simplest cases at AdS tree level were computed long ago using SUSY constraints and perturbation methods (e.g., \cite{Eden:2000bk,Arutyunov:2002fh,Arutyunov:2017dti,Arutyunov:2018tvn}). With the help of Mellin space techniques \cite{Mack:2009mi,Mack:2009gy,Penedones:2010ue,Fitzpatrick:2011ia} a unified formula was discovered for arbitrary four CPOs \cite{Rastelli:2016nze,Rastelli:2017udc}. Because of the essential complexity of bulk perturbation such computation barely continued its success to loop levels. 

Recently there arose a new approach from the boundary side, where analytic conformal bootstrap methods are available (e.g., \cite{Komargodski:2012ek,Fitzpatrick:2012yx,Alday:2016njk,Caron-Huot:2017vep,Carmi:2019cub,Mazac:2019shk,Penedones:2019tng,Caron-Huot:2020adz}, a recent review is available in \cite{Bissi:2022mrs}). In brief the idea is to rebuild the correlator from its Lorentzian singularities. When applied to theories with a weakly coupled bulk dual, this induces recursive relations among CFT data at different orders of bulk perturbation \cite{Aharony:2016dwx}. Specializing to $\mathcal{N}=4$ SYM such analysis succeeded in generating new results at one loop \cite{Alday:2017xua,Aprile:2017bgs,Alday:2017vkk,Aprile:2017qoy,Aprile:2019rep,Alday:2019nin}.

In this paper we report the computation for the four-point correlator of the simplest CPOs, the stress tensor multiplet, at bulk two loops. 

\section{Preliminaries}

\subsection{Structure of $\langle\mathcal{O}_2\mathcal{O}_2\mathcal{O}_2\mathcal{O}_2\rangle$}

In $\mathcal{N}=4$ SYM with gauge group $\mathrm{SU}(N)$, the CPOs $\mathcal{O}_p$ are half-BPS states with protected conformal dimension $p$ (for some integer $p\geq2$) that transform as a Lorentz scalar and in the $[0,p,0]$ (i.e., traceless symmetric) representation of the $\mathrm{SU}(4)$ R-symmetry. We normalize them such that
\begin{equation}
\langle\mathcal{O}_p(x_1,y_1)\mathcal{O}_p(x_2,y_2)\rangle=\left(\frac{y_{12}^2}{x_{12}^2}\right)^p,
\end{equation}
where $x_{ij}^2=(x_i-x_j)^2$ and $y_{ij}^2=y_i\cdot y_j$. $x^\mu$ is the spacetime coordinates. The 6d null vector $y^a$ is a convenient way to take care of the R-symmetry representation. Among these operators the simplest case $p=2$ is the stress tensor multiplet and is dual to the graviton in $\mathrm{AdS}_5\times\mathrm{S}^5$. 

As usual we define the crossratios
\begin{equation}
    \begin{split}
u\equiv\,&\frac{x_{12}^2x_{34}^2}{x_{13}^2x_{24}^2}=z\bar{z},\quad
v\equiv\frac{x_{14}^2x_{23}^2}{x_{13}^2x_{24}^2}=(1-z)(1-\bar{z}),\\
&\frac{y_{12}^2y_{34}^2}{y_{13}^2y_{24}^2}=\alpha\bar{\alpha},\quad\!\qquad
\frac{y_{14}^2y_{23}^2}{y_{13}^2y_{24}^2}=(1-\alpha)(1-\bar{\alpha}),
\end{split}
\end{equation}
By superconformal Ward identities the correlator $\langle2222\rangle\equiv\langle\mathcal{O}_2(x_1,y_1)\cdots\mathcal{O}_2(x_4,y_4)\rangle$ can be written as \cite{Eden:2000bk}
\begin{equation}\label{eq:partialnonrenormalization}
\begin{split}
&\left(\frac{y_{12}^2y_{34}^2}{x_{12}^2x_{34}^2}\right)^{-2}\langle2222\rangle\equiv\mathcal{G}
=\mathcal{G}^{\rm free}+\mathcal{I}\,\mathcal{H},
\end{split}
\end{equation}
where $\mathcal{I}$ is a simple factor
\begin{equation}
    \mathcal{I}=\frac{(z-\alpha)(z-\bar{\alpha})(\bar{z}-\alpha)(\bar{z}-\bar{\alpha})}{(z\bar{z}\alpha\bar{\alpha})^2}.
\end{equation}
The free correlator $\mathcal{G}^{\rm free}$ is completely determined by Wick contraction and is independent of the 't Hooft coupling $\lambda$ (its explicit expression can be found in, e.g., (2.11) of \cite{Aprile:2017bgs}). So non-trivial dynamics of $\langle2222\rangle$ is fully encoded in $\mathcal{H}$, which is usually called the reduced correlator. Consider the supergravity limit $N\to\infty$ and $\lambda\to\infty$, where we can take double expansion of $\mathcal{H}$ with respect to small $1/\sqrt{\lambda}$ and $1/c$, $c=\frac{N^2-1}{4}$ being the central charge. $1/\sqrt{\lambda}$ controls the stringy corrections to the bulk effective action, which are beyond our current scope of investigation, and so in this paper we only focus on the supergravity sector, at order $\lambda^0$ \footnote{Due to this ignorance of stringy corrections our supergravity result is subject to counterterms whose coefficients are not yet determined. This will be further discussed later in the paper.}. The $1/c$ expansion of this sector take the form
\begin{equation}\label{eq:Hcexpand}
\mathcal{H}\vert_{\rm sugra}=c^{-1}\mathcal{H}^{(1)}+c^{-2}\mathcal{H}^{(2)}+c^{-3}\mathcal{H}^{(3)}+\cdots.
\end{equation}
This expansion has a clean bulk interpretation in terms of perturbation in Newton's constant. $\mathcal{H}^{(1)}$ corresponds to the tree-level scattering, and was computed long ago in \cite{DHoker:1999mic,Arutyunov:2000py}. $\mathcal{H}^{(2)}$ corresponds to the one-loop scattering, and its position space result was recently obtained using analytic bootstrap on the CFT side \cite{Aprile:2017bgs}. Our target in this paper is to extend the position space computation and solve $\mathcal{H}^{(3)}$, which describes the graviton scattering at two loops in this theory.

Because our computation relies on the used of CFT data, it is helpful to quickly review the conformal block expansion of $\langle2222\rangle$. The superconformal symmetry of SYM nicely organizes its spectrum into various supermultiplets. Among these there are the protected multiplets (including the half-BPS and the semi-short multiplets) and the long multiplets. Correspondingly the correlator $\mathcal{G}$ can as well be decomposed as
\begin{equation}\label{eq:Gshortlong}
    \mathcal{G}=\mathcal{G}_{\rm protected}+\mathcal{I}\mathcal{H}_{\rm long}.
\end{equation}
The contribution from the protected multiplets $\mathcal{G}_{\rm protected}$ are free from quantum corrections and its expression was determined in \cite{Doobary:2015gia,Aprile:2017bgs}. The remaining part is then identical to a summation over long multiplets only. Because the long multiplets appearing in the OPE of $\mathcal{O}_2\times\mathcal{O}_2$ can only R-symmetry representation $[0,0,0]$, whose corresponding superconformal blocks are explicitly
\begin{equation}\label{eq:longblock}
    A_{\tau,\ell}^{[0,0,0]}(z,\bar{z},\alpha,\bar{\alpha})=\mathcal{I}g_{\tau+4,\ell}(z,\bar{z}),
\end{equation}
where $g_{\tau,\ell}$ denotes the ordinary conformal block
\begin{align}
    \label{eq:Gblock}g_{\tau,\ell}&=\frac{z\bar{z}}{\bar{z}-z}\left(k_{\frac{\tau-2}{2}}(z)k_{\frac{\tau+2\ell}{2}}(\bar{z})-(z\leftrightarrow\bar{z})\right),\\
\label{eq:2F1}k_h(z)&=z^h{}_2F_1(h,h,2h,z).
\end{align}
The shift of twist in $g$ relative to $A$ in \eqref{eq:longblock} is consistent with the non-trivial conformal weights carried by the factor $\mathcal{I}$. This indicates that the quantity $\mathcal{H}_{\rm long}$ in \eqref{eq:Gshortlong} only depends on the position space cross-ratios and it receives a decomposition onto the ordinary blocks as
\begin{align}\label{eq:HOPE}
\mathcal{H}_{\rm long}(z,\bar{z})&=\sum_{i}a_ig_{\tau_i+4,\ell_i}(z,\bar{z}),
\end{align}
where the index $i$ scans over the spectrum of the long multiplets in the theory.

In the supergravity limit, the supergravity piece of $\mathcal{H}_{\rm long}$ can be expanded in analogy to \eqref{eq:Hcexpand}
\begin{equation}
    \mathcal{H}_{\rm long}\vert_{\rm sugra}=\mathcal{H}_{\rm long}^{(0)}+c^{-1}\mathcal{H}_{\rm long}^{(1)}+c^{-2}\mathcal{H}_{\rm long}^{(2)}+c^{-3}\mathcal{H}_{\rm long}^{(3)}+\cdots.
\end{equation}
Note that in the case of $\langle 2222\rangle$ both $\mathcal{G}_{\rm free}$ and $\mathcal{G}_{\rm protected}$ are non-vanishing only up to order $c^{-1}$. This implies the identification
\begin{equation}
    \mathcal{H}^{(k)}=\mathcal{H}_{\rm long}^{(k)},\qquad k\geq2.
\end{equation}
In other words, from the boundary point of view the graviton scattering at one loop and higher only receives contributions from long multiplets. Because in the supergravity limit all the multiplets in correspondence to stringy states become infinitely heavy and get decoupled, these remaining long multiplets have their twists at most finite at the leading order and are known to be in general multi-trace operators.

To see the structure of $\mathcal{H}^{(k)}$ at loop levels more clearly, consider the expansion of twists and OPE coefficients in \eqref{eq:HOPE} with respect to small $1/c$ (we suppress stringy corrections)
\begin{align}
\label{eq:twistexpansion}\tau_i&=\tau_i^{(0)}+c^{-1}\gamma_i^{(1)}+c^{-2}\gamma_i^{(2)}+\cdots,\\
\label{eq:coefexpansion}a_i&=a_i^{(0)}+c^{-1}a_i^{(1)}+c^{-2}a_i^{(2)}+\cdots.
\end{align}
Operators with $a_i^{(0)}\neq0$ already show up in the mean field theory, and are the double-trace operators of the generic form $[\mathcal{O}_{q_1}\mathcal{O}_{q_2}]_{n,\ell}\equiv\mathcal{O}_{q_1}\square^n\partial^\ell\mathcal{O}_{q_2}$, with classical twist $\tau^{(0)}=q_1+q_2+2n$. So in the OPE $\mathcal{O}_2\times\mathcal{O}_2$ all the classical twists are even, with the smallest value at $\tau^{(0)}=4$. From \eqref{eq:Gblock} and \eqref{eq:2F1} one observes that (in s-channel) $\mathcal{H}^{(k)}$ contains terms with the factor $\log{u}$ up to $\log^k{u}$. Terms with the maximal power $k$ are called the leading log terms, which are

\begin{equation}\label{eq:leadingloggeneric}
\mathcal{H}^{(k)}\big|_{\log^k {u}}=\!\sum_{\tau^{(0)},\ell}\!\frac{\langle a^{(0)}(\gamma^{(1)})^k\rangle_{\tau^{(0)},\ell} }{p!\,2^p}g_{\tau^{(0)}+4,\ell}(z,\zb).
\end{equation}
The coefficients only depend on $a^{(0)}$ and $\gamma^{(1)}$, so this part solely consists of double-trace operators, and can be recursively determined once these data at the disconnected and the tree level are available. At $\tau^{(0)}=4$ it is literally the product $a^{(0)}(\gamma^{(1)})^k$, as there is a unique operator $\mathcal{O}_2\mathcal{O}_2$. At higher $\tau^{(0)}$ there are multiple choices of $\{q_1,q_2,n\}$ for the same classical twist, and the angle bracket indicates certain average due to degeneracy among these operators. For more details of this degeneracy, see e.g., \cite{Aprile:2017xsp,Aprile:2018efk}. More generally, at higher loops other terms with $\log^{\geq2}u$ can be recursively constructed from lower-loop data in analogous ways. One example is shown in \eqref{eq:twist4}, which will be used as one of the inputs to our computation. However, very often the determination of these subleading log terms requires data of higher-trace operators, which are not available by considering 4-point half-BPS correlators only.

The significance of $\mathcal{H}^{(k)}|_{\log^k{u}}$ can already be seen at one loop, where it fully determines the behavior of $\langle2222\rangle$ around its Lorentzian singularities, which further reconstructs the correlator \cite{Alday:2017xua,Aprile:2017bgs}. Starting at two loops, while these leading log terms still plays an important role \cite{Bissi:2020woe}, the Lorentzian singularities also relies on the subleading log terms as well. Due to the lack of information about higher-trace operators as mentioned above, one cannot expects to compute $\mathcal{H}^{(3)}$ as straightforwardly as at one loop. Therefore we need to look for alternative sources of constraints for this quantity.

\subsection{Hidden Symmetries}

A useful hint already resides in the structure of the leading log terms. It was noted in \cite{Caron-Huot:2018kta} that the expansion \eqref{eq:leadingloggeneric} can in fact resum into a very intuitive analytic expression
\begin{equation}\label{eq:leadinglogrelationsharp}
\mathcal{H}^{(k)}\big|_{\log^k {u}}=\left[\Delta^{(8)}\right]^{k-1}\mathcal{F}^{(k)}(z,\bar{z}).
\end{equation}
$\Delta^{(8)}$ is an eighth-order differential operator 
\begin{align}
\Delta^{(8)} = \frac{z\zb}{\zb-z} D_z(D_z-2)D_{\zb}(D_{\zb}-2) \frac{\zb-z}{z\zb}
\end{align}
acting in s-channel, where $D_z=z^2\partial_z (1-z)\partial_z$. \eqref{eq:leadinglogrelationsharp} is a consequence of a conjectured 10d conformal symmetry observed from the pattern of tree-level data, where the degeneracy in $\langle a^{0}(\gamma^{(1)})^k\rangle$ is interpreted in terms of decomposition of $\mathrm{SO}(10,2)$ blocks into those of its $\mathrm{SO}(4,2)\times\mathrm{SO}(6)$ subgroup.  
The function $\mathcal{F}^{(k)}(z,\bar{z})$ is built out of 10d blocks, which takes the generic form
\begin{equation}\label{eq:leadingF}
    \mathcal{F}^{(k)}(z,\bar{z})=\sum_{|\vec{a}|=0}^k\frac{p_{\vec{a}}(z,\bar{z})}{(\bar{z}-z)^7}G(\vec{a};z)+(z\leftrightarrow\bar{z}).
\end{equation}
$G(\vec{a};z)$ denotes the multiple polylogarithm (MPL), labeled by a weight vector $\vec{a}$. In this case, components of $\vec{a}$ take values in $\{0,1\}$. $p_{\vec{a}}(z,\bar{z})$ is some polynomial of $z$ and $\bar{z}$, of degree $7$ in each. Finally, the summation over MPLs is carried up to weight $k$. Detailed expressions were worked out in \cite{Caron-Huot:2018kta}.

Interestingly, \cite{Aprile:2019rep} discovered that similar action of $\Delta^{(8)}$ as in \eqref{eq:leadinglogrelationsharp} simplifies the structure of $\langle2222\rangle$ at one-loop as well, with a slight modification \footnote{The relation between our $\mathcal{H}^{(2)}$ here and their $\mathcal{H}^{(2)}_{2222}$ is $\mathcal{H}^{(2)} =u^2 \mathcal{H}^{(2)}_{2222}/4 $.}
\begin{align}\label{eq:H2result}
\mathcal{H}^{(2)} = \Delta^{(8)} \mathcal{L}^{(2)} + \frac{1}{4} \mathcal{H}^{(1)},
\end{align}
where
\begin{align}\label{eq:L2}
\mathcal{L}^{(2)} = \sum_{|\vec{a}|+|\vec{a}'|=0}^4 \frac{p_{\vec{a},\vec{a}'}(z,\zb)}{(\zb-z)^7} G(\vec{a};z)G(\vec{a}';\zb).
\end{align}
Like $p_{\vec{a}}$, the polynomial $p_{\vec{a},\vec{a}'}(z,\zb)$ is again of degree $7$ in each variable. Now the summation is performed over all possible combinations $G(\vec{a};z)G(\vec{a}';\zb)$ of total weight up to $4$. Note this maximal weight $4=2+2$ is identical to the weight of the leading log terms including $\log^2 {u}$.

This is a huge simplification as compared to the expression of $\mathcal{H}^{(2)}$ itself. It turns out $\mathcal{L}^{(2)}$ can in fact be fully bootstraped by generic consistency constratints on $\mathcal{L}^{(2)}$ and $\mathcal{H}^{(2)}$, without even knowing the leading log terms \cite{Aprile:2019rep}! This strongly suggests a much wider use of $\Delta^{(8)}$.

\section{Ansatz for $\mathcal{H}^{(3)}$}

Based on the above discussions we draw the following ansatz for the two-loop correlator of $\langle2222\rangle$
\begin{align}\label{eq:H3ansatz}
\mathcal{H}^{(3)} = \left[ \Delta^{(8)} \right]^2 \mathcal{L}^{(3)} + \kappa_2 \mathcal{H}^{(2)} + \kappa_1 \mathcal{H}^{(1)},
\end{align}
where $\kappa_1$ and $\kappa_2$ are two parameters to be determined later on. Similar to \eqref{eq:L2} the unknown function $\mathcal{L}^{(3)}$ admits a decomposition onto MPLs, now up to weight $6$. Here we can set up the ansatz in a more informed manner. The full correlator $\mathcal{H}^{(3)}$ is expected to be single-valued on the Euclidean slice $\bar{z}=z^*$, with singularities at $0,1,\infty$. When multiplied by $(z-\bar{z})^4/(z\bar{z})^4$ it becomes invariant under an $\mathrm{S}_3\times \mathbb{Z}_2$ permutation group. The $\mathbb{Z}_2$ associates to the exchange of variables $z\leftrightarrow\bar{z}$. The $\mathrm{S}_3$ permutes $0,1,\infty$, and consists of the following operations
\begin{equation}\label{eq:s3trans}
    \begin{split}
        &\rho(123):\,z\mapsto z,\quad\quad\quad\;\;\,\rho(132):\,z\mapsto\bar{z}/(\bar{z}-1),\\
        &\rho(213):\,z\mapsto1-\bar{z},\quad\quad\rho(231):\,z\mapsto1/(1-z),\\
        &\rho(312):\,z\mapsto1-1/z,\quad\rho(321):\,z\mapsto1/\bar{z},
    \end{split}
\end{equation}
(and $\bar{z}$ transforming in the conjugate way), which account for the full crossing symmetries of the correlator. Therefore it is natural to choose a basis for single-valued MPLs (SVMPLs) where each element falls into a specific representation of $\mathrm{S}_3\times\mathbb{Z}_2$. We label them as
\begin{equation}
    G^{\rm SV}_{w,r,i}(z,\bar{z}),
\end{equation}
where $w$ denotes the uniform transcendental weight, and $r$ chooses the six different $\mathrm{S}_3\times \mathbb{Z}_2$ representations $\{\mathbf{1}^\pm,\bar{\mathbf{1}}^\pm,\mathbf{2}^\pm\}$ ($\pm$ stands for symmetric/anti-symmetric representation of $\mathbb{Z}_2$). The third index $i$ takes care of linearly independent SVMPLs with the same $w$ and $r$. $\mathrm{S}_3$ acts on these functions as
\begin{align}
    \rho(abc)\circ G^{\rm SV}_{w,r,i}&=\mathbf{R}_{(abc)}^rG^{\rm SV}_{w,r,i},
\end{align}
where $\rho(abc)\circ $ means to change the variables according to \eqref{eq:s3trans}, and $\mathbf{R}_{(abc)}^r$ is the matrix for the element $(abc)$ in representation $r$. 
Explicitly, we have $\mathbf{R}_{(abc)}^{\mathbf{1}^\pm}\equiv1$.  $\mathbf{R}_{(abc)}^{\bar{\mathbf{1}}^\pm}$  is $+1$ for even $(abc)$ permutations and $-1$ otherwise. $G^{\rm SV}_{w,\mathbf{2}^{\pm},i}$ is an array of two basis functions, and the corresponding linear transformation can be realized in various ways. The choice that we make is 
\begin{equation}
    \begin{split}
        &\mathbf{R}_{(123)}^{\mathbf{2}^\pm}=\left(\begin{matrix}1&0\\0&1\end{matrix}\right),\quad\;\;
        \mathbf{R}_{(132)}^{\mathbf{2}^\pm}=\left(\begin{matrix}1&-1\\0&-1\end{matrix}\right),\\
        &\mathbf{R}_{(213)}^{\mathbf{2}^\pm}=\left(\begin{matrix}0&1\\1&0\end{matrix}\right),\quad\;\;
        \mathbf{R}_{(231)}^{\mathbf{2}^\pm}=\left(\begin{matrix}0&-1\\1&-1\end{matrix}\right),\\
        &\mathbf{R}_{(312)}^{\mathbf{2}^\pm}=\left(\begin{matrix}-1&1\\-1&0\end{matrix}\right),\quad
        \mathbf{R}_{(321)}^{\mathbf{2}^\pm}=\left(\begin{matrix}-1&0\\-1&1\end{matrix}\right).
    \end{split}
\end{equation}

MPLs are accompanied by a useful tool called \emph{symbol} (see appendix). Provided a finite set of allowed symbol alphabets, one can construct such basis of SVMPLs for any cutoff in the maximal transcendental weight. Up to one loop the alphabets are observed to be $\{z,\bar{z},1-z,1-\bar{z}\}$, in accordance with the singularities of the correlator at $0$, $1$ and $\infty$. We expect this to still be valid at two loops, except that a fifth letter $z-\bar{z}$ is also present (note this pattern already appear in the denominator of \eqref{eq:L2}). The basis construction can be systematized following, e.g., the algorithm introduced in  \cite{Chavez:2012kn}\footnote{In that paper the authors constructed the SVMPL basis functions by taking care of the $\mathbb{Z}_2$ group only. We follow their strategy but consider $\mathrm{S}_3$ representations as well.}. Without the letter $z-\bar{z}$, the counting of independent representations for each $w$ and $r$ are as follows
\begin{center}
    \begin{tabular}{c|cccccc}
         \toprule
         \;\diagbox[height=0.8cm,width=1.6cm]{$w$}{$r$}\; & $\;\mathbf{1}^+\;$ & $\;\mathbf{1}^-\;$ & $\;\bar{\mathbf{1}}^+\;$ & $\;\bar{\mathbf{1}}^-\;$ & $\;\mathbf{2}^+\;$ & $\;\mathbf{2}^-\;$ \\
         \midrule
         $0$ & $1$ & $0$ & $0$ & $0$ & $0$ & $0$ \\
         $1$ & $0$ & $0$ & $0$ & $0$ & $1$ & $0$ \\
         $2$ & $2$ & $1$ & $0$ & $0$ & $1$ & $0$ \\
         $3$ & $2$ & $0$ & $1$ & $0$ & $3$ & $1$ \\
         $4$ & $5$ & $3$ & $1$ & $0$ & $5$ & $2$ \\
         $5$ & $7$ & $3$ & $4$ & $2$ & $11$ & $5$ \\
         $6$ & $15$ & $10$ & $6$ & $3$ & $20$ & $12$ \\
         \bottomrule
    \end{tabular}
\end{center}
For instance, up to $w=2$ we have 
\begin{equation}\label{eq:basisexamples}
    \begin{split}
        &G_{0,\mathbf{1}^+}^{\rm SV}\!=\!1,\;\;
        G_{1,\mathbf{2}^+}^{\rm SV}\!=\!(\log{u},\log{v}),\;\;
        G_{2,\mathbf{1}^+,1}^{\rm SV}\!=\!\zeta_2,\\ &G_{2,\mathbf{1}^+,2}^{\rm SV}=\log^2 {u}-\log{u}\log{v}+\log^2 {v},\\
        &G_{2,\mathbf{1}^-}^{\rm SV}\!=\!2\mathrm{Li}_2(z)-2\mathrm{Li}_2(\bar{z})+\log{u}\log\frac{1-z}{1-\bar{z}},\\
        &G_{2,\mathbf{2}^+}^{\rm SV}\!\!=\!\left(\!\frac{\log^2\! {u}\!-\!2\log{u}\log{v}}{2},\frac{\log^2\! {v}\!-\!2\log{u}\log{v}}{2}\!\right)\!.
    \end{split}
\end{equation}
Here we give an explicit example where the SVMPL of give $w$ and $r$ is not unique, in which case there are multiple basis functions $G_{2,\mathbf{1}^+,i}^{\rm SV}$ (with $i=1,2$). In such situation one can of course make arbitrary choice of basis functions (e.g., $G_{2,\mathbf{1}^+,1}^{\rm SV}+G_{2,\mathbf{1}^+,2}^{\rm SV}$ and $G_{2,\mathbf{1}^+,1}^{\rm SV}-G_{2,\mathbf{1}^+,2}^{\rm SV}$ here), but the dimension of the function space is invariant.

When the letter $z-\bar{z}$ is also included, starting at weight $3$ there are additional independent functions. In particular at weight $3$ there is a unique $\mathbf{1}^-$ function, which we denote as $Q_3$ \footnote{The $Q_3$ function we construct here is to be distinguished from that defined in \cite{Chavez:2012kn}. We require this function to be in the $\mathbf{1}$ representation of $\mathrm{S}_3$.}. There are even more functions at higher weights but it turns out only $Q_3$ is needed (intuition about these extra functions is discussed in the appendix). Hence in total there are 189 independent SVMPLs in our basis, and $\mathcal{L}^{(3)}$ has the expansion
\begin{align}
\label{eq:L3ansatz}\mathcal{L}^{(3)} &= \sum_{w=0}^6\sum_{r,i}\frac{\omega^{w,r}_{i}(z,\zb)}{(\zb-z)^7} G^{\rm SV}_{w,r,i}(z,\bar{z}),\\
\label{eq:omega}\omega^{w,r}_{i}&=\sum_{j,k=0}^7c^{w,r}_{i,j,k}z^j\bar{z}^k,\quad c^{w,r^\pm}_{i,j,k}=\mp c^{w,r^\pm}_{i,k,j}.
\end{align}
The undetermined coefficients $c^{w,r}_{i,j,k}$ are rational. The condition \eqref{eq:omega} is to ensure that $\mathcal{L}^{(3)}$ is a $\mathbb{Z}_2$ invariant. The power of denominator in \eqref{eq:L3ansatz}, $7$, is exactly the same as that at one-loop level \eqref{eq:L2} and that in the leading log terms at arbitrary loops \eqref{eq:leadingF}. One may be curious about taking a larger power (for instance $9$) to begin with, and so there will be even more unknown variables. With the constraints to be described later on, in our computation we find such enlarged ansatz leads to the same result.

\section{Bootstrap}

We now set out to constrain the above ansatz.
$\mathcal{L}^{(3)}$ is not a full correlator, but is expected to admit a decomposition into s-channel blocks. Therefore a set of generic properties can be directly imposed at the level of $\mathcal{L}^{(3)}$, as listed below:
\begin{enumerate}
	\item In Euclidean region $\mathcal{L}^{(3)}$ should be finite at $z = \zb$.
	\item As a sum of s-channel blocks with identical external operators, exchanging operator 1 and 2 should leave $\mathcal{L}^{(3)}$ unchanged. Explicitly we have
	\begin{equation}
	    \mathcal{L}^{(3)}(z,\zb)=\mathcal{L}^{(3)}\left(\frac{\zb}{\zb-1},\frac{z}{z-1}\right).
	\end{equation}
\end{enumerate}
The known leading log expression \eqref{eq:leadinglogrelationsharp} provides an important constraint on $\mathcal{H}^{(3)}$. In fact it can be imposed directly on $\mathcal{L}^{(3)}$, as $\Delta^{(8)}$ acting on $\log^k {u}$ definitely lowers its power. This leads to a theory-specific condition:
\begin{enumerate}
    \item[A.] When expanding $\mathcal{L}^{(3)}$ in the s-channel, all the $\log^k {u}$ terms with $k>3$ have to vanish. And the $\log^3 {u}$ terms of $\mathcal{L}^{(3)}$ should match known data. By comparing \eqref{eq:L3ansatz} and \eqref{eq:leadinglogrelationsharp} we have
    \begin{equation}\label{eq:leadinglogmatch}
        \mathcal{L}^{(3)}(z,\bar{z})\big|_{\log^3 {u}}=\mathcal{F}^{(3)}(z,\bar{z}),
    \end{equation}
    where the detailed expression of $\mathcal{F}^{(3)}$ can be found in Appendix C of \cite{Caron-Huot:2018kta}.
\end{enumerate}

After solving the above constraints we pass on to the full correlator $\mathcal{H}^{(3)}$ via the action of $[\Delta^{(8)}]^2$. Firstly there is one last generic CFT constraint:
\begin{enumerate}
	\item[3.] $\mathcal{H}^{(3)}$ should respect the full crossing symmetries. With the help of the $\mathrm{S}_3\times\mathbb{Z}_2$ SVMPL basis, this is equivalent to requiring that, when decomposing the correlator as
	\begin{equation}
	    \frac{(z-\bar{z})^4}{z^4\bar{z}^4}\mathcal{H}^{(3)}=\!\!\!\!\sum_{w,r,i}\Omega^{w,r}_{i}(z,\bar{z})G^{\rm SV}_{w,r,i}(z,\bar{z}),
	\end{equation}
	the rational coefficient functions $\Omega^{w,r}_{i}(z,\bar{z})$ transform in a way such that each term on RHS is an $\mathrm{S}_3$ invariant, i.e.,
	\begin{align}\label{eq:coefomega}
	    \rho(abc)\circ \Omega^{w,r}_{i}&=\Omega^{w,r}_{i}(\mathbf{R}_{(abc)}^r)^{-1}.
	\end{align}
\end{enumerate}
Then there are more conditions specific to $\mathcal{N}=4$ SYM, mainly regarding its spectrum:
\begin{enumerate}
	\item[B.] The tree-level correlator $\mathcal{H}^{(1)}$ contains twist below the double-twist threshold $4$ in its decomposition into ordinary conformal blocks, which is manifested in that $\mathcal{H}^{(1)}$ has poles at $z=1$ and at $\bar{z}=1$. As pointed out before, for $\langle2222\rangle$ $\mathcal{H}^{(k\geq2)}$ consists of only long super-multiplets (with $\tau^{(0)}\geq4$) and so is free of these poles. Therefore in our ansatz \eqref{eq:H3ansatz} $\left[\Delta^{(8)}\right]^2\mathcal{L}^{(3)}$ expects to produce exactly the same poles (which is indeed true) in order to cancel out the ones in $\mathcal{H}^{(1)}$. This condition fixes $\kappa_1$ to $-\frac{1}{16}$.
	\item[C.] Coefficients of the subleading $\log^2 {u}$ terms in $\mathcal{H}^{(3)}$ take the schematic form $\langle a^{(1)}(\gamma^{(1)})^2+2a^{(0)}\gamma^{(1)}\gamma^{(2)}\rangle$. In general they include contributions from long operators other than the double-trace ones $[\mathcal{O}\mathcal{O}]$. Due to the lack of knowledge about these extra operators these coefficients cannot be recursively determined in the fashion of leading log coefficients \eqref{eq:leadingloggeneric}. 
	However, twist $4$ is a lucky exception, where only $[\mathcal{O}_2\mathcal{O}_2]_{0,\ell}$ contribute. There is even no need to solve any operator mixing and these coefficients are easily determined as
	\begin{equation}\label{eq:twist4}
	\begin{split}
	    \langle a^{(1)}(\gamma^{(1)})^2+2a^{(0)}\gamma^{(1)}\gamma^{(2)}\rangle_{4,\ell}=&\frac{2\langle a^{(0)}\gamma^{(1)}\rangle_{4,\ell}\langle a^{(1)}\gamma^{(1)}+a^{(0)}\gamma^{(2)}\rangle_{4,\ell}}{\langle a^{(0)}\rangle_{4,\ell}}\\&-\frac{\langle a^{(0)}\gamma^{(1)}\rangle_{4,\ell}^2\langle a^{(1)}\rangle_{4,\ell}}{\langle a^{(0)}\rangle_{4,\ell}^2},
	\end{split}
	\end{equation}
	where the data shown on RHS are all available from $\langle 2222\rangle$ at lower orders. This fixes $\kappa_2$ to $\frac{5}{4}$.
	\item[D.] It is generally expected that in the bulk-point limit $\langle2222\rangle$ reduces to the flat-space four-graviton scattering amplitude in 10d \cite{Heemskerk:2009pn,Gary:2009ae,Maldacena:2015iua}. \cite{Alday:2017vkk} proposed a simpler connection between the two sides at loop level, by comparing the discontinuities. Specifically, the leading divergence of $\mathrm{dDisc}\,\mathcal{H}(z^{\circlearrowright},\zb)$ at the bulk point limit ($z=\bar{z}+2\epsilon\bar{z}\sqrt{1-\bar{z}}$, $\epsilon\to0$ after continuing $z$ around $z=0$ clock-wisely) should match the discontinuity of the scattering amplitude $\mathcal{A}$ across the t-channel cut. At two loops we have
	\begin{align}
    &(\zb-z)^{23}\text{dDisc}\, \mathcal{H}^{(3)}(z^{\circlearrowright},\zb)\big\vert_{z\to\zb} 
    =\frac{\Gamma(22)(1-\zb)^{11}\zb^{24}}{8\pi^8} \text{Disc}_{x>1}\mathcal{A}^{(2)}(x)\big\vert_{x=1/\zb},
    \end{align}
	where the two-loop supergravity amplitude $\mathcal{A}^{(2)}(x)$ has been computed in \cite{Bissi:2020woe}. This matching can also be done even before the action of $[\Delta^{(8)}]^2$. See discussions in Section 2.4 of \cite{Bissi:2020woe}.
	\item[E.] By construction it is not guaranteed that the ansatz produces OPE data with sufficient analyticity in spin. At two loops it is expect that the OPE coefficient are analytic functions of spin $\ell$ down to $\ell=2$ at $\log^2 u$, and to $\ell=6$ at $\log^1 u$ and $\log^0 u$, while the non-analyticity at lower spins are closely tied to the allowed contourterm ambiguities at two loops. To ensure sufficient analyticity we impose that the ansatz is self-consistent under CFT dispersion relation, up to these ambiguities. In our computation this constraint is implemented using the so-called Lorentzian inversion formula \cite{Caron-Huot:2017vep,Simmons-Duffin:2017nub,Alday:2017vkk,Caron-Huot:2018kta}, and the details are presented in the appendix.

\end{enumerate}

Note that after the action of $[\Delta^{(8)}]^2$ the number of unsolved $c^{w,r}_{i,j,k}$ that actually show up in $\mathcal{H}^{(3)}$ is reduced. This is because $[\Delta^{(8)}]^2$ has a non-trivial kernel, which generates no physical effects at all. Apart from these disappearing variables there are two types of remaining degrees of freedom at the end of the above computation. The first type can be identified as ambiguities due to counterterms, which are discussed in the next section. 

The second type is genuinely tied to supergravity, which has support on infinitely many spins. This type contains \textit{only one} remaining variable (let us call it $\mathcal{X}$). We are not aware of direct physical constraints to determine this variable. However, there are considerations regarding the mathematical structure of the result. 

Note that at the end of the above computation the function space in which the correlator lives greatly shrinks, by the fact that a large portion of basis functions $G^{\rm SV}$ completely disappear \footnote{In $(w,r)$ sectors with degeneracy this observation relies on a further proper choice of basis functions. For instance, in the $(2,\mathbf{1}^+)$ sector, with the basis shown in \eqref{eq:basisexamples} we observe $G_{2,\mathbf{1}^+,1}^{\rm SV}$ disappears at the end.}. By curiosity one can try further reducing the function space by tuning the remaining variables. Remarkably, it turns out the only possibility is to fix $\mathcal{X}$ to a \textit{unique} value! To be explicit, this value kills one $
G_{5,\bar{\mathbf{1}}^-}^{\rm SV}$ and one $G_{4,\mathbf{2}^+}^{\rm SV}$ basis elements. Furthermore, while $\mathcal{L}^{(3)}$ does not need to respect the full crossing symmetry in general, we find it is possible to consistently do so, and this fixes $\mathcal{X}$ to exactly the same value as above. These are strong hints that this choice should be special from the physical perspective as well, hence we conjecture that it leads to the correct supergravity result at two loops.

\section{Results}

Loop-level supergravity computations suffer from counterterms of the schematic form $\partial^{2k}\mathcal{R}^4$. These vertices can build up tree-level and one-loop diagrams, which we denote as $\mathcal{C}_{\text{tree},i}$ and $\mathcal{C}_{\text{one-loop},q}$, $i$ and $q$ labeling different diagrams. These diagrams are responsible for the cancellation of UV divergence in the two-loop correlator of supergravity, at the cost of introducing ambiguities unless the full string theory is considered. 

The tree-level diagrams $\mathcal{C}_{\text{tree},i}$ are easily described in Mellin space \cite{Mack:2009mi,Mack:2009gy,Penedones:2010ue}
\begin{equation}
    \mathcal{C}_{\text{tree},i} = \int_{-i\infty}^{i\infty} \frac{\dd s\, \dd t}{(2\pi i)^2}u^{\frac{s}{2}+2}v^{\frac{t}{2}-2}\mathcal{M}_{\text{tree},i}\, \Gamma\left(2-\frac{s}{2}\right)^2\Gamma\left(2-\frac{t}{2}\right)^2\Gamma\left(2-\frac{\Tilde{u}}{2}\right)^2,
\end{equation}
with $s+t+\Tilde{u}=4$. The corresponding Mellin amplitudes $\mathcal{M}_{\text{tree},i}$ are just symmetric polynomials in $s,t,\Tilde{u}$ \cite{Alday:2014tsa,Alday:2018pdi}
\begin{align}
    \mathcal{M}_{\text{tree},1}&= 1,\\
    \mathcal{M}_{\text{tree},2} &= s^2+t^2+\Tilde{u}^2,\\
    \mathcal{M}_{\text{tree},3} &= s t \Tilde{u},\\
    \mathcal{M}_{\text{tree},4} &= (s^2+t^2+\Tilde{u}^2)^2,\\
    \mathcal{M}_{\text{tree},5} &= s t \Tilde{u}(s^2+t^2+\Tilde{u}^2),\\
    \mathcal{M}_{\text{tree},6} &= s^2 t^2 \Tilde{u}^2.
\end{align}
For those who are not familiar with Mellin space method, we also offer the results in position space
\begin{align}
    \mathcal{C}_{\text{tree},1} &= 4u^4(1+u\partial_u)(2+u\partial_u)(3+u\partial_u)\partial_u^3 \Phi(u,v),\\
    \mathcal{C}_{\text{tree},i} &= \widehat{\mathcal{M}}_{\text{tree},i}\, \mathcal{C}_{\text{tree},1},
\end{align}
where
\begin{equation}
    \Phi(u,v) = \frac{1}{z-\zb}\left( \text{Li}_2(z) - \text{Li}_2(\zb) + \log(z\zb)\log\frac{1-z}{1-\zb} \right),
\end{equation}
and $\widehat{\mathcal{M}}_{\text{tree},i}$ are differential operators obtained by replacing $s,t,\Tilde{u}$ in $\mathcal{M}_{\text{tree},i}$ with
\begin{align}
    \hat{s} &= -4+2u\partial_u,\\
    \hat{t} &= 4+2v\partial_v,\\
    \hat{u} &= 4-2u\partial_u-2v\partial_v.
\end{align}
Among these tree-level ambiguities only $\mathcal{C}_{\rm tree,i\leq5}$ appear in our computation, because $\mathcal{C}_{\rm tree,6}$ has a power of $z-\bar{z}$ in the denominator that exceeds our ansatz, while the others can all be put into the pattern as an action of $(\Delta^{(8)})^2$ or just come from the leftover ambiguity in the one-loop reduced correlator $\mathcal{H}^{(2)}$.

The one-loop diagrams $\mathcal{C}_{\text{one-loop},i}$ no longer take such a simple form in Mellin space. The origin of these diagrams is further described in Appendix \ref{sec:zzbar}. Like the tree-level contact diagrams, they have the characteristic feature that their leading log OPE coefficients have finite support on spin \cite{Alday:2018pdi,Drummond:2019hel}. With this feature these diagrams can be bootstrapped individually using similar ansatz. The allowed solutions can be classified according to the power of $z-\bar{z}$ in the denominator in its leading log ($\log^2u$) part, which is $7+2q$ ($q=0,1,\ldots$) \cite{Alday:2018pdi}. By this power counting the ambiguities that can possibly show up in our ansatz corresponds to the solutions with $q$ up to $8$. We list these nine solutions in the ancillary file. It turns out that in order to fit the $(\Delta^{(8)})^2$ structure in \eqref{eq:H3ansatz} they are further restricted to two possible linear combinations, which are responsible for the remaining one-loop ambiguities left over in our computation.

We have checked that the remaining degrees of freedom from our computation (except for $\mathcal{X}$) belong to the counterterm ambiguities described above. 
Nevertheless, because we do not study string corrections in this paper, all the above ambiguities can in principle enter the complete supergravity result, including those that are not captured by our ansatz. We therefore express the final result of this bootstrap as a special solution plus all such counterterm ambiguities
\begin{equation}\label{eq:H3result}
\boxed{
\begin{split}
\mathcal{H}^{(3)} =& \left[ \Delta^{(8)} \right]^2 \mathcal{L}^{(3)} + \frac{5}{4} \mathcal{H}^{(2)} - \frac{1}{16} \mathcal{H}^{(1)}\\
&+ \sum_{q=0}^\infty b_{1,q} \mathcal{C}_{\text{1-loop},q} + \sum_{i=1}^6 b_{0,i} \mathcal{C}_{\text{tree},i}.
\end{split}
}
\end{equation}
Notice that $\mathcal{C}_{\text{1-loop},q}$ may potentially affect $\log^1$ and $\log^0$ data for all spin $\ell$. The coefficients $b_{1,q}$ and $b_{0,i}$ can be determined only when the full string theory corrections to the correlator are considered. We leave them for future work.

In $\mathcal{H}^{(3)}$ each coefficient function $\Omega^{w,r}_{i}(z,\bar{z})$ forms a representation of $\mathrm{S}_3\times\mathbb{Z}_2$ according to \eqref{eq:coefomega}. This motivate us to define a basic building block
\begin{equation}
    B_{d,k}(z,\bar{z})=\frac{(1+u+v)^{d-3k}(uv)^k}{(z-\bar{z})^d},\quad d,k\in\mathbb{N},
\end{equation}
and then in each representation $r$ and given maximal power $d_{\rm max}$ of $(z-\bar{z})$ we construct a basis $\mathcal{B}^{r}_{d_{\rm max}}$ for $\Omega$
\begin{align}
    \mathcal{B}^{\mathbf{1}^+}_{d_{\rm max}}&=\{B_{d,k}\;|\;d\text{ even},d\leq d_{\rm max},k\leq\lfloor d/3\rfloor\},\\
    \mathcal{B}^{\mathbf{1}^-}_{d_{\rm max}}&=\{B_{d,k}\;|\;d\text{ odd},d\leq d_{\rm max},k\leq\lfloor d/3\rfloor\},\\
    \mathcal{B}^{\bar{\mathbf{1}}^\pm}_{d_{\rm max}}&=\frac{(1-u)(1-v)(u-v)}{(z-\bar{z})^3}\mathcal{B}^{\mathbf{1}^\mp}_{d_{\rm max}-3},\\
    \mathcal{B}^{\mathbf{2}^\pm}_{d_{\rm max}}&=\frac{(1-2u+v,1+u-2v)}{(z-\bar{z})}\mathcal{B}^{\mathbf{1}^\mp}_{d_{\rm max}-1}\nonumber\\
    &\;\quad\bigcup\frac{(1-2u^2+v^2,1+u^2-2v^2)}{(z-\bar{z})^2}\mathcal{B}^{\mathbf{1}^\pm}_{d_{\rm max}-2}.
\end{align}
For our study of $\mathcal{H}^{(3)}$, $d_{\rm max}=19$. Then the correlator can be expressed as
\begin{equation}
    \frac{(z-\bar{z})^4}{z^4\bar{z}^4}\mathcal{H}^{(3)}=\sum_{r}\mathcal{B}^{r}_{19}\mathcal{C}^{r}\mathcal{G}^{\rm SV}_{r},
\end{equation}
and the dynamics of the theory is encoded in the coefficient matrices $\mathcal{C}^{r}$. $\mathcal{G}^{\rm SV}_{r}$ denotes the set of SVMPL basis for the representation $r$. 

Comparing to the ansatz, we end up with a much smaller set of SVMPL basis. Apart from $Q_3$, the counting of the basis in each sector reduces as
\begin{center}
    \begin{tabular}{c|cccccc}
         \toprule
         \;\diagbox[height=0.8cm,width=1.6cm]{$w$}{$r$}\; & $\;\mathbf{1}^+\;$ & $\;\mathbf{1}^-\;$ & $\;\bar{\mathbf{1}}^+\;$ & $\;\bar{\mathbf{1}}^-\;$ & $\;\mathbf{2}^+\;$ & $\;\mathbf{2}^-\;$ \\
         \midrule
         $0$ & $1$ & $0$ & $0$ & $0$ & $0$ & $0$ \\
         $1$ & $0$ & $0$ & $0$ & $0$ & $1$ & $0$ \\
         $2$ & ${\color{black!40}2\to}1$ & $1$ & $0$ & $0$ & $1$ & $0$ \\
         $3$ & ${\color{black!40}2\to}1$ & $1$ & $1$ & $0$ & ${\color{black!40}3\to}2$ & $1$ \\
         $4$ & ${\color{black!40}5\to}2$ & ${\color{black!40}3\to}2$ & $1$ & $0$ & ${\color{black!40}5\to}1$ & $2$ \\
         $5$ & ${\color{black!40}7\to}1$ & ${\color{black!40}3\to}1$ & ${\color{black!40}4\to}1$ & ${\color{black!40}2\to}1$ & ${\color{black!40}11\to}3$ & ${\color{black!40}5\to}3$ \\
         $6$ & ${\color{black!40}15\to}0$ & ${\color{black!40}10\to}2$ & ${\color{black!40}6\to}0$ & ${\color{black!40}3\to}1$ & ${\color{black!40}20\to}0$ & ${\color{black!40}12\to}2$ \\
         \bottomrule
    \end{tabular}
\end{center}
The $\mathcal{C}^{r}$ matrices are still too huge to directly fit into this paper, and we provide them in the ancillary file on arXiv. Because the constraints on $\mathcal{X}$ is conjectural, in that file we also present the coefficients of $\mathcal{X}$ before this last step for readers' reference (we have shifted the definition of this last variable such that $\mathcal{X}=0$ corresponds to our supergravity result).

\section{Outlook}

In this paper we bootstraped the simplest CPO correlator in $\mathcal{N}=4$ SYM at AdS two loops. Our computation is based on an ansatz that generalizes a structure observed at one loop which was inspired by a conjectured 10d conformal symmetry. Very interestingly, after exploiting all possible physical constraints available to us, there arises a unique solution (up to contourterm ambiguities). We expect that this result may inspire detailed studies of graviton dynamics and scattering of other particles at two loops and at even higher loops. Combined with string corrections that fix the counterterms this will provide data for unprotected operators in SYM at strong coupling with higher precision. Some of the future explorations are commented below.

Our supergravity result can be expressed in terms of repeated action of a remarkable differential operator $\Delta^{(8)}$. In particular, if we plug in the one-loop result \eqref{eq:H2result} into \eqref{eq:H3result} we obtain
\begin{align}
\mathcal{H}^{(3)} = \left[ \Delta^{(8)} \right]^2 \mathcal{L}^{(3)} +\frac{5}{4} \Delta^{(8)}\mathcal{L}^{(2)} + \frac{1}{4} \mathcal{L}^{(1)},
\end{align}
plus additional counterterms, where $\mathcal{L}^{(1)}=\mathcal{H}^{(1)}$. The coefficients above are both very simple rationals, which puts strong confidence on the validity of this result as well as the structure of the ansatz. It is then natural to write the correlator $\langle2222\rangle$ at arbitrary higher loops as
\begin{equation}
    \mathcal{H}^{(k)}=\sum_{j=1}^k \kappa^{(k)}_{j}\left[\Delta^{(8)}\right]^{j-1}\mathcal{L}^{(j)},
\end{equation}
where $\kappa^{(k)}_{k}=1$. Then at each order $k$ the problem reduces to solving a single function $\mathcal{L}^{(k)}$ together with the $\kappa^{(k)}_{j<k}$ coefficients. From the structure of the known leading log terms $\mathcal{L}^{(j)}$ need to contain SVMPLs of weights up to $2k$, with coefficients of the same structure as in \eqref{eq:L3ansatz}\eqref{eq:omega}.  

Moreover, the conjectured hidden symmetries in \cite{Caron-Huot:2018kta} also implies that at tree-level other CPO correlators $\langle pqrs\rangle$ (after subtracting a free part) can be extracted from $\langle2222\rangle$ by another differential operator $\mathcal{D}_{pqrs}$
\begin{align}
\label{eq:treehidden}
\mathcal{H}^{(1)}_{pqrs} - \mathcal{H}^{(1),\text{free}}_{pqrs}= \mathcal{D}_{pqrs} \mathcal{H}^{(1)}.
\end{align}
The leading log terms at higher loops again enjoy the precise relation
\begin{align}
\label{eq:llhidden}
\mathcal{H}^{(k)}_{pqrs}\Big\vert_{\log^k {u}} = \left[ \Delta^{(8)} \right]^{k-1} \mathcal{D}_{pqrs} \mathcal{F}^{(k)}(z,\zb),
\end{align}
where $\Delta^{(8)}$ now takes a more general form, which together with $\mathcal{D}_{pqrs}$ can be found in \cite{Caron-Huot:2018kta}. It was observed in \cite{Aprile:2019rep} that one-loop correlators $\mathcal{H}^{(2)}_{22pp}$ admit representations of the form \eqref{eq:H2result} as well, although the non-derivative term is no longer simply proportional to $\mathcal{H}^{(1)}_{22pp}$. It will be interesting to investigate whether this may suggest similar method to bootstrap generic $\mathcal{H}^{(3)}_{pqrs}$.

Finally, observe that the number of independent SVMPLs is greatly reduced in the final result as compared to those in the ansatz. In many categories with fixed $w$ and $r$ there is even a unique function remaining. It would be nice to see whether these functions individually have any physical significance.

\acknowledgments

The authors would like to thank Yunfeng Jiang, Song He, Paul Heslop, Lilin Yang, Yang Zhang, Xinan Zhou, Huaxing Zhu for useful discussions. ZH and EYY are supported by National Science Foundation of China under Grant No.~12175197 and Grand No.~12147103. EYY is also supported by National Science Foundation of China under Grant No.~11935013, and by the Fundamental Research Funds for the Chinese Central Universities under Grant No.~226-2022-00216.

\appendix

\section{Multiple Polylogarithms}

Multiple polylogarithms (MPLs) can be defined in terms of nested integrals \cite{Duhr:2019tlz}
\begin{align}
G(z)=&1,\nonumber\\
G(\vec{a};z) \equiv& G(a_1,...,a_n;z)= \int_{0}^{z} \frac{\dd t}{t-a_1} G(a_2,...,a_n;t).\nonumber
\end{align}
Especially when $\vec{a}=(0,\dots,0)$, we define
\begin{align}
G(\underbrace{0,\dots,0}_{n};z)=\frac{1}{n!}\log{z}^n.\nonumber
\end{align}
The length $|\vec{a}|$ of vector $\vec{a}$ is called the weight of $G(\vec{a};z)$. 
$G$ is naturally accompanied by a mathematical entity called \textit{symbol}, $\mathcal{S}[G]$, whose advantages include encoding information about their singularities and simplifying identities among MPLs. $\mathcal{S}[G]$ can be obtained by applying differentiation iteratively. If
\begin{equation*}
    \dd G = \sum_i G'_i\  \dd \log R_i,
\end{equation*}
where $R_i$s are algebraic functions, then
\begin{equation*}
    \mathcal{S}[G] = \sum_i S[G'_i] \otimes  R_i.
\end{equation*}
As a simple example, $\mathcal{S}[G(0,1;z)]\equiv\mathcal{S}[-\mathrm{Li}_2(z)] = (1-z) \otimes  z$. Expression in each entry of the $\otimes$ product is called a \emph{symbol letter}, and the collection of all letters the \emph{alphabet} of the symbol. Our computation of both MPLs and their symbols utilizes the PolyLogTools Mathematica package introduced in \cite{Duhr:2019tlz}.

\section{About Extra MPLs with Letter $z-\bar{z}$}\label{sec:zzbar}

There are several intuitions about why $Q_3$ has to be included in the computation at two loops. On the one hand, one may try to carry out the same bootstrap program as above but without $Q_3$. Then in the comparison at the bulk-point limit the ansatz fails to match the flat-space result. In fact, it manages to cover the contribution from plannar diagrams in the supergravity amplitude, but that from the non-plannar diagrams goes beyond.

On the other hand, when viewing in the full context of string theory with $\lambda$ expansion, the one-loop supergravity correlator receives a string correction from $\mathcal{R}^4$ residing at the $\lambda^0$ order (e.g., see \cite{Chester:2019pvm,Chester:2020dja}). By unitarity it together with the tree-level supergravity result recursively determines a one-loop counterterm to the two-loop correlator, which is again at the same $\lambda^0$ order as supergravity. Moreover, by origin this correction has to be proportional to the $\lambda^{-3/2}$ correction to the one-loop correlator, and the latter was computed in \cite{Drummond:2019hel} and was observed to contain $Q_3$ \footnote{Again, the $Q_3$ function in that paper differs from ours by MPLs without the symbol letter $z-\bar{z}$.}. Because in our setup the computation is subject to ambiguities regarding counterterms, there is no reason that this function is well separated from the supergravity.

One may also wonder about higher-weight single-valued MPLs with letter $z-\zb$. We found that including such MPLs with weight 4 leads to the same result.

\section{Constraints from Lorentzian Inversion Formula}

As mentioned previously the last step of our computation constrains the ansatz using Lorentzian inversion formula \cite{Caron-Huot:2017vep,Simmons-Duffin:2017nub}. As with other types of dispersion relations, this formula reconstructs (data of) the entire correlator from its singularities in the Lorentzian region. The core object here is the double discontinuity
\begin{equation}
    \mathrm{dDisc}\,\mathcal{H}(z,\bar{z})=\mathcal{H}(z,\bar{z})-\frac{1}{2}\mathcal{H}^{\circlearrowleft}(z,\bar{z})-\frac{1}{2}\mathcal{H}^{\circlearrowright}(z,\bar{z}),
\end{equation}
where $\circlearrowleft$ and $\circlearrowright$ means to analytically continue $\bar{z}$ around $\bar{z}=1$ counter-clockwisely and clockwisely respectively. 
Specific to the computation of $\langle2222\rangle$ this formula reads \cite{Alday:2017vkk,Caron-Huot:2018kta}\newline
\begin{equation}\label{eq:Lorentzianinversion}
    C(\tau,\ell)\!=\!\kappa_{\bar{h}}\!\int_0^1\frac{\mathrm{d}z}{z^2}\frac{\mathrm{d}\bar{z}}{\bar{z}^2}k_{1-h}(z)k_{\bar{h}}(\bar{z})\mathrm{dDisc}\left[\frac{\bar{z}-z}{\bar{z}z}\mathcal{H}\right],
\end{equation}
where $\kappa_{\bar{h}}=\frac{(1+(-1)^\ell)\Gamma(\hb)^4}{4\pi^2\Gamma(2\hb)\Gamma(2\hb-1)}$, $h=\frac{\tau+2}{2}$ and $\bar{h}=\frac{\tau+2\ell+4}{2}$. It encodes the twists and the coefficients of blocks in the s-channel OPE in the form
\begin{equation}\label{eq:ctl}
    C(\tau,\ell)=\sum_{i}\frac{a_i}{\tau-\tau_i},
\end{equation}
where the summation is over different operators. Taking $1/c$ expansion \eqref{eq:twistexpansion}\eqref{eq:coefexpansion} into consideration RHS of \eqref{eq:ctl} further expands out, e.g., at order $c^{-3}$
\begin{equation}\label{eq:ctlexpansion}
    \begin{split}
         C(\tau,\ell)\supset
        \frac{1}{c^3}\sum_{\tau^{(0)}}&\left ( \frac{\langle a^{(3)}\rangle}{\tau-\tau^{(0)}}+\frac{\langle a^{(2)}\gamma^{(1)}+a^{(1)}\gamma^{(2)}+a^{(0)}\gamma^{(3)}\rangle}{(\tau-\tau^{(0)})^2} \right. \\
        &\left. \quad +\frac{\langle a^{(1)}(\gamma^{(1)})^2+2a^{(0)}\gamma^{(1)}\gamma^{(2)}\rangle}{(\tau-\tau^{(0)})^3}+\frac{\langle a^{(0)}(\gamma^{(1)})^3\rangle}{(\tau-\tau^{(0)})^4}\right).
    \end{split}
\end{equation}
These $\tau$ poles in $C(\tau,\ell)$ originate from $z$ integrals in \eqref{eq:Lorentzianinversion}. To see this, we first expand the integrand of RHS at $z=0$. Since $k_{1-h}(z)\sim z^{1-h}(1+\frac{1-h}{2}z+\cdots)$, the integrals to be done are
\begin{equation}\label{eq:zint}
\begin{split}
    \hspace{-2em}\int_0^1\! \frac{\dd z}{z^2} z^{1-h+n}\log^k z = \frac{(-1)^{k+1}k!}{(h\!-\!n)^{k+1}} \sim \frac{1}{(\tau\!+\!2\!-\!2n)^{k+1}}.
    \end{split}
\end{equation}
Notice that only $\log^{k\geq k_0} z$ terms contribute to $(k_0+1)$-th poles. For the computation of $\bar{z}$ integrals, see Appendix A in \cite{Alday:2017zzv}.

The data collected from the various coefficients in \eqref{eq:ctlexpansion} directly show up in the block expansion of $\mathcal{H}^{(3)}$ as well (as can be easily worked out by taking $1/c$ expansion of \eqref{eq:HOPE})
\begin{equation}\label{eq:Hresum}
\begin{split}
\mathcal{H}^{(3)} =&\sum_{\tau^{(0)},\ell} \left( \langle a^{(3)}\rangle\ g_{\tau^{(0)},\ell} + \langle a^{(2)}\gamma^{(1)}+a^{(1)}\gamma^{(2)}+a^{(0)}\gamma^{(3)}\rangle\, \partial_{\tau^{(0)}}g_{\tau^{(0)},\ell}\right.\\&\left.+ \langle a^{(1)}(\gamma^{(1)})^2+2a^{(0)}\gamma^{(1)}\gamma^{(2)}\rangle \frac{1}{2!}\partial^2_{\tau^{(0)}}\,g_{\tau^{(0)},\ell} + \langle a^{(0)}(\gamma^{(1)})^3\rangle  \frac{1}{3!}\partial^3_{\tau^{(0)}}\,g_{\tau^{(0)},\ell}\right).
\end{split}
\end{equation}
Once we obtain these data from the Lorentzian inversion formula, we plug them into the above expression \eqref{eq:Hresum} to reproduce $\mathcal{H}^{(3)}$ and compare with the ansatz. This serves as a consistency constraint at each 
$\log u$ orders (note that in \eqref{eq:Hresum} $\log u$ can come out from $\partial^k_{\tau^{(0)}}\, g_{\tau^{(0)},\ell}$, which tops up to $\log^k u$). For example at the leading log order
\begin{equation}\label{eq:resum}
    \sum_{\tau^{(0)},\ell} \frac{\langle a^{(0)}(\gamma^{(1)})^3\rangle}{3!\,2^3}\, g_{\tau^{(0)}+4,\ell}(z,\bar{z}) = \mathcal{H}^{(3)}\big\vert_{\log^3 {u}}.
\end{equation}
However, \eqref{eq:resum} is a double infinite sum of two-variable functions, which is hard to handle. A better way is to consider the series expansion of \eqref{eq:resum} up to some order $\mathcal{O}(z^{h_\text{max}})$. Since $g_{\tau^{(0)}+4,\ell}(z,\bar{z})\sim\mathcal{O}(z^{\frac{\tau^{(0)}+4}{2}})$, the $\tau^{(0)}$ sum can thus be truncated to a finite sum
\begin{equation}\label{eq:resumct}
\begin{split}
    \sum_{\tau^{(0)} = 4}^{\scriptsize 2h_\text{max}-4} \sum_{\ell=0}^\infty \frac{\langle a^{(0)}(\gamma^{(1)})^3\rangle}{3!\,2^3}\, g_{\tau^{(0)}+4,\ell}-\mathcal{H}^{(3)}\big\vert_{\log^3 {u}} = \mathcal{O}(z^{h_\text{max}+1}).
\end{split}
\end{equation}
For the remaining infinite sum over $\ell$, by \eqref{eq:Gblock} we only need to work out for each $\tau^{(0)}$
\begin{equation}\label{eq:resumell}
    \sum_{\ell=0}^\infty \frac{\langle a^{(0)}(\gamma^{(1)})^3\rangle}{3!\,2^3}\, k_{2+\frac{\tau^{(0)}}{2}+\ell}(\bar{z}).
\end{equation}
This infinite sum is supposed to be a linear combination of MPLs with alphabet $\{\bar{z},1-\bar{z}\}$, where coefficients are rational functions with denominator $\bar{z}^{\frac{\tau^{(0)}}{2}}$. The analytic result can be easily worked out by setting up an ansatz and matching the Taylor series around $\bar{z}=0$ with \eqref{eq:resumell}.

During the actual computation, since the leading log terms were already determined by the matching \eqref{eq:leadinglogmatch} in previous steps, \eqref{eq:resumct} itself only provides a consistency check. However, we observe that non-trivial constraints enter in the comparison at subsequent $\log u$ terms. We exhaust all possible constraints of this type by gradually lifting the twist cutoff $h_{\rm max}$ till no more constraints further appear.

Subtlety comes when analyticity in spin can not go down to $\ell=0$ \cite{Caron-Huot:2017vep}. For example the $\log^2 {u}$ data $\langle a^{(1)}(\gamma^{(1)})^2+2a^{(0)}\gamma^{(1)}\gamma^{(2)}\rangle$ extracted from $C(\tau,\ell)$ at $\ell=0$ may not be the true OPE data of $\mathcal{H}^{(3)}$. In this case, we set $\ell=0$ data as unknown variables when summing with conformal blocks. The same applies to $l=0,2,4$ data in $\log^1 u$ and $\log^0 u$. This strategy takes care of possible counterterm ambiguities appearing in the two-loop calculations.

\bibliographystyle{JHEP}
\bibliography{N4SYMc3}

\end{document}